\documentclass[proceedings,article,submit,moreauthors]{Definitions/mdpi}

\usepackage{layouts}

\def\doirms{$\mathrm{DOI}_{\mathrm{RMS}}$}

\firstpage{1}
\pubvolume{1}
\issuenum{1}
\articlenumber{0}
\pubyear{2026}
\copyrightyear{2026}
\datereceived{}
\daterevised{}
\dateaccepted{}
\datepublished{}

\Title{Pulse-shape discrimination with machine learning for CZT detectors at the DA$\Phi$NE beam test facility}

\providecommand{\orcidlink}[1]{\href{https://orcid.org/#1}{\orcidicon}}

\Author{%
Simone Manti $^{1,}$*\orcidlink{0000-0003-3770-0863},
Francesco Artibani $^{1,2}$\orcidlink{0009-0000-8905-3165},
Leonardo Abbene $^{3,1}$\orcidlink{0000-0001-9633-6606},
Massimiliano Bazzi $^{1}$\orcidlink{0000-0002-1699-7138},
Manuele Bettelli $^{4}$\orcidlink{0000-0003-4062-3782},
Giacomo Borghi $^{5,6}$\orcidlink{0000-0001-8488-4728},
Damir Bosnar $^{7}$\orcidlink{0000-0003-4784-393X},
Mario Bragadireanu $^{8}$\orcidlink{0009-0001-5217-6003},
Antonino Buttacavoli $^{3,1}$\orcidlink{0000-0002-7188-3651},
Mario Carminati $^{5,6}$\orcidlink{0000-0001-9734-3007},
Alberto Clozza $^{1}$\orcidlink{0000-0003-2133-1725},
Francesco Clozza $^{1,9}$\orcidlink{0009-0002-3298-0624},
Luca De Paolis $^{1}$\orcidlink{0000-0002-4203-9902},
Raffaele Del Grande $^{10,1}$\orcidlink{0000-0002-7599-2716},
Kamil Dulski $^{1,11,12}$\orcidlink{0000-0002-4093-8162},
Carlo Fiorini $^{5,6}$\orcidlink{0000-0002-1157-0143},
Ivica Fri\v{s}\v{c}i\'c $^{7}$\orcidlink{0000-0002-4743-0572},
Gaetano Gerardi $^{3}$\orcidlink{0009-0005-2542-1850},
Carlo Guaraldo $^{1}$\orcidlink{0000-0002-8923-3438},
Mihai Iliescu $^{1}$\orcidlink{0009-0003-3859-5679},
Masa Iwasaki $^{13}$\orcidlink{0000-0002-3460-9469},
Alexander Khreptak $^{11,12,1}$\orcidlink{0000-0002-9482-9770},
Johan Marton $^{14,15}$\orcidlink{0009-0003-1912-285X},
Pawel Moskal $^{11,12}$\orcidlink{0000-0001-5644-5963},
Hiroaki Ohnishi $^{16}$\orcidlink{0000-0001-9427-1984},
Kristian Piscicchia $^{17,1}$\orcidlink{0000-0001-6879-452X},
Fabio Principato $^{3,1}$\orcidlink{0000-0003-2787-0877},
Alessandro Scordo $^{1}$\orcidlink{0000-0002-7703-7050},
Francesco Sgaramella $^{1}$\orcidlink{0000-0002-0011-8864},
Michał Silarski $^{11}$\orcidlink{0000-0003-2206-0963},
Diana Sirghi $^{17,1,8}$\orcidlink{0009-0002-7486-025X},
Florin Sirghi $^{1,8}$\orcidlink{0000-0002-6143-3200},
Magdalena Skurzok $^{11,12,1}$\orcidlink{0000-0002-4794-5154},
Antonio Spallone $^{1}$\orcidlink{0009-0000-2111-8014},
Kairo Toho $^{16,1}$\orcidlink{0009-0001-3245-1418},
Oton Vazquez Doce $^{1}$\orcidlink{0000-0001-6459-8134},
Andrea Zappettini $^{4}$\orcidlink{0000-0002-6916-2716},
Johann Zmeskal $^{14}$\orcidlink{0000-0003-0815-0639} and
Catalina Curceanu $^{1,8}$\orcidlink{0000-0002-1990-0127}}

\longauthorlist{yes}

\AuthorNames{Simone Manti, Francesco Artibani, Leonardo Abbene, Massimiliano Bazzi, Manuele Bettelli, Giacomo Borghi, Damir Bosnar, Mario Bragadireanu, Antonino Buttacavoli, Mario Carminati, Alberto Clozza, Francesco Clozza, Luca De Paolis, Raffaele Del Grande, Kamil Dulski, Carlo Fiorini, Ivica Friščić, Gaetano Gerardi, Carlo Guaraldo, Mihai Iliescu, Masa Iwasaki, Alexander Khreptak, Johan Marton, Pawel Moskal, Hiroaki Ohnishi, Kristian Piscicchia, Fabio Principato, Alessandro Scordo, Francesco Sgaramella, Michał Silarski, Diana Sirghi, Florin Sirghi, Magdalena Skurzok, Antonio Spallone, Kairo Toho, Oton Vazquez Doce, Andrea Zappettini, Johann Zmeskal and Catalina Curceanu}

\address{%
$^{1}$ \quad Laboratori Nazionali di Frascati INFN, Frascati, Italy\\
$^{2}$ \quad Università degli Studi di Roma Tre, Dipartimento di Fisica, Roma, Italy\\
$^{3}$ \quad Department of Physics and Chemistry (DiFC), Emilio Segrè, University of Palermo, Palermo, Italy\\
$^{4}$ \quad Istituto Materiali per l'Elettronica e il Magnetismo, Consiglio Nazionale delle Ricerche, Parco Area delle Scienze 37/A, 43124 Parma, Italy\\
$^{5}$ \quad Politecnico di Milano, Dipartimento di Elettronica, Informazione e Bioingegneria, Milano, Italy\\
$^{6}$ \quad INFN Sezione di Milano, Milano, Italy\\
$^{7}$ \quad Department of Physics, Faculty of Science, University of Zagreb, Zagreb, Croatia\\
$^{8}$ \quad IFIN-HH, Institutul Național Pentru Fizică și Inginerie Nucleară Horia Hulubei, 30 Reactorului, 077125, Măgurele, Romania\\
$^{9}$ \quad Università degli Studi di Roma Tor Vergata, Dipartimento di Fisica, Roma, Italy\\
$^{10}$ \quad Faculty of Nuclear Sciences and Physical Engineering, Czech Technical University in Prague, Břehovà 7, 115 19, Prague, Czech Republic\\
$^{11}$ \quad Faculty of Physics, Astronomy, and Applied Computer Science, Jagiellonian University, Kraków, Poland\\
$^{12}$ \quad Center for Theranostics, Jagiellonian University, Krakow, Poland\\
$^{13}$ \quad RIKEN, Tokyo, Japan\\
$^{14}$ \quad Stefan Meyer Institute for Subatomic Physics, Vienna, Austria\\
$^{15}$ \quad Atominstitut, Technische Universität Wien, Stadionallee 2, 1020 Vienna, Austria\\
$^{16}$ \quad Research Center for Accelerator and Radioisotope Science (RARiS), Tohoku University, Sendai, Japan\\
$^{17}$ \quad Centro Ricerche Enrico Fermi, Museo Storico della Fisica e Centro Studi e Ricerche ``Enrico Fermi'', Roma, Italy}

\corres{Correspondence: Simone.Manti@lnf.infn.it}

\abstract{Cadmium zinc telluride (CZT) detectors offer versatility, operational simplicity, and room-temperature X- and gamma-ray spectroscopy, making them attractive for collider applications, yet their use under high-flux conditions remains limited. Here, we present a preliminary feature-based pulse-shape analysis employing machine learning, on data acquired with a quasi-hemispherical CZT detector at the DA$\Phi$NE beam test facility of the National Laboratory of Frascati of INFN. A 300-MeV electron beam impinging on a lead target produced characteristic Pb X-rays together with a broad background extending up to the electron–positron annihilation region. Physically motivated temporal and morphological features were extracted from the recorded waveforms and used to distinguish nominal photon-like pulses from anomalous events. An XGBoost classifier trained and validated on 10,000 labeled waveforms achieved an accuracy of approximately 97\%, with most of its classification performance reached using only a few hundred labeled examples. The trained model was applied to more than 700,000 events, substantially reducing the spectral continuum and coincidence peaks, while preserving the characteristic Pb X-ray lines up to the 511-keV annihilation peak. These preliminary results demonstrate the potential of machine-learning-assisted pulse-shape discrimination for improving CZT spectroscopy in collider environments.}


\let\linenumbers\relax
\begin{document}
%\AtBeginDocument{\nolinenumbers}
%
% \nolinenumbers
%\printinunitsof{in}\prntlen{\textwidth}
%
\section{Introduction}
%
%----------X- and gamma-ray spectroscopy requirements----------
High-resolution X- and gamma-ray spectroscopy is an essential tool in nuclear and particle physics \cite{Eberth_2008_GeLiDetectors,Gotta_2004_PrecisionSpectroscopy}, astrophysics \cite{Harrison_2013_NUCLEARSPECTROSCOPIC}, medical imaging \cite{Taguchi_2013_Vision20}, and plasma diagnostics \cite{Renner_2019_ChallengesXray}. Many of these applications require compact detection systems capable of operating over a broad energy range and under different experimental conditions. In particular, experiments characterized by high particle rates or intense electromagnetic backgrounds require detectors combining good spectroscopic performance, high detection efficiency, and stable operation over extended acquisition periods.\newline\noindent
%
% Add references on applications of X- and gamma-ray spectroscopy.
%
%----------CZT detectors and room-temperature operation----------
Cadmium zinc telluride (CZT) detectors represent an attractive solution for room-temperature photon spectroscopy \cite{Iniewski_2024_CdTeCdZnTe}. Their high effective atomic number and density provide good photon-stopping efficiency, while their wide band gap allows operation without cryogenic cooling. These properties enable the development of compact detection systems with good spectroscopic performance, achieving an energy resolution of a few \% over energies ranging from a few keV to hundreds of keV \cite{Mele_2023_AdvancesHighEnergyResolution}, extending the possibility to detect from X-rays to gamma rays \cite{Eisen_1998_CdTeCdZnTe}. However, variations in charge transport and collection inside the crystal can produce different pulse morphologies, incomplete charge collection, and low-energy tails. Consequently, the recorded waveform contains information about the interaction and charge-collection processes that is not fully represented by the reconstructed pulse amplitude alone.\newline\noindent
%
%----------Operation in accelerator environments----------
The operation of CZT detectors in accelerator environments is particularly challenging because of the intense and complex electromagnetic background \cite{Abbene_2020_RoomtemperatureXray}. In addition to single photon interactions, the acquired data may contain multiple energy depositions, pile-up events, charged-particle interactions, electronic artifacts, accidental triggers, and saturated pulses. Some of these contributions overlap in energy with the signals of interest and cannot therefore be efficiently rejected using only the reconstructed energy. The identification of additional features sensitive to the event topology is consequently important for improving peak visibility and the signal-to-background ratio.\newline\noindent
%
% Add references on semiconductor detectors operated in high-rate or
% accelerator environments.
%
%----------CZT detectors in SIDDHARTA-2----------
Within the SIDDHARTA-2 experiment \cite{Sirghi_2024_SIDDHARTA2Apparatusb} at the DA$\Phi$NE collider \cite{Milardi_2018_PreparationActivitya,Milardi_2021_DAFNECommissioning,Milardi_2024_DAFNEOperation}, CZT detectors are being investigated for the spectroscopy of intermediate-mass kaonic atoms. The characteristic transitions of these systems extend above the energy range primarily covered by the silicon drift detectors employed for light kaonic atoms. Previous measurements at DA$\Phi$NE demonstrated the feasibility of operating CZT detectors in a collider environment \cite{Scordo_2024_CdZnTeDetectors} and allowed several characteristic X-ray transitions to be observed in kaonic fluorine and copper \cite{Artibani_2026_KaonicCopper}. Nevertheless, the residual beam-induced background limits the visibility of weak spectral features and represents an important challenge for future precision measurements, as planned by the EXKALIBUR project \cite{Manti_2025_EXKALIBURKaonic}.\newline\noindent
%
% Add references on SIDDHARTA-2 and previous CZT measurements at DA$\Phi$NE.
%
% Add references on the BTF facility and the experimental campaign.
%
%----------Pulse-shape analysis----------
Pulse-shape analysis (PSA) provides a possible approach for distinguishing photon-like events from background contributions \cite{McGrath_2010_DetectingMultihit,Nakhostin_2013_ApplicationPulseshape,Fritts_2014_PulseshapeDiscrimination}. The temporal development of the detector signal reflects the charge-collection process and may reveal multiple interactions, pile-up, anomalous rise profiles, saturation, or the absence of a physical pulse. Physically interpretable features, including the rise time, the number of peaks in the waveform derivative, and proxies for the interaction depth and charge-collection dynamics, can therefore be used to characterize the recorded events. Multivariate and machine learning (ML) methods can exploit correlations among these features, potentially improving event classification compared to selection criteria based solely on individual features \cite{Holl_2019_DeepLearninga,Manti_2026_MachinelearningenhancedEvent}.\newline\noindent
%
% Add references on pulse-shape analysis and feature-based event selection.
%
%----------Contribution of this work----------
In this work, we present a preliminary PSA of data acquired with a quasi-hemispherical CZT detector with the DA$\Phi$NE beam test facility (BTF) at the National Laboratory of Frascati of INFN (INFN-LNF). Representative waveform morphologies are identified and characterized using a limited set of physically motivated pulse-shape features. The distributions of these features are investigated, and their capability to reject non-photon-like events is evaluated by comparing the amplitude spectrum before and after event selection. Particular attention is devoted to the visibility of the characteristic Pb X-ray lines and of the 511-keV annihilation peak. The purpose of this study is to demonstrate the feasibility of an interpretable waveform-based background-suppression strategy for CZT detectors operating in an accelerator environment.\newline\noindent
%
%----------Paper organization----------
The paper is structured as follows. Section~2 describes the experimental setup, the CZT detection system, and the waveform-processing and feature-extraction procedures. Section~3 presents the feature distributions, the preliminary classification results, and the effect of the event selection on the measured spectrum. Finally, Section~4 summarizes the main conclusions and outlines future developments.
\section{Experimental setup and methods}
\begin{figure}[h!]
    \centering
    \includegraphics[width=\linewidth]{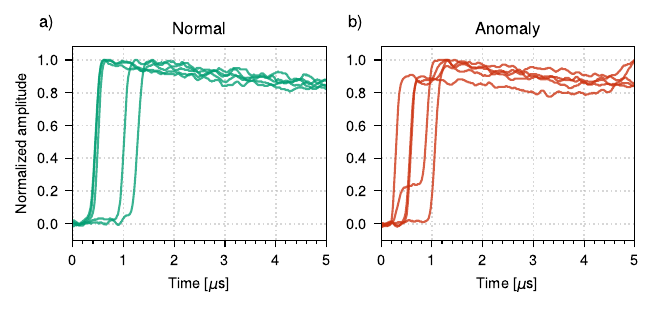}
    \caption{Examples of waveforms categorized as normal in panel (a) and events assigned to the anomaly class in panel (b).}
    \label{fig:waveforms}
\end{figure}

%
%----------BTF and beam conditions----------
The data used in this work were collected during a dedicated measurement campaign in 2026 at the DA$\Phi$NE BTF of the INFN-LNF, where the BTF provides pulsed electron or positron beams with adjustable energy and intensity \cite{Foggetta_2021_ExtendedOperative}. For the present measurements, a 300-MeV electron beam impinged on a 1-mm-thick lead target. The beam–target interactions produced characteristic Pb X-rays, accompanied by bremsstrahlung and secondary particles. The resulting spectrum covered the characteristic Pb fluorescence lines (L$\alpha$, K$\alpha$ and K$\beta$) and the higher-energy continuum, including the 511-keV electron–positron annihilation line. This configuration provided a controlled accelerator environment for investigating pulse-shape signatures associated with beam-induced backgrounds relevant to, although not fully representative of, those encountered at the interaction point of the DA$\Phi$NE collider, as in the SIDDHARTA-2 experiment.\newline\noindent
\begin{figure}[h!]
    \centering
    \includegraphics[width=\linewidth]{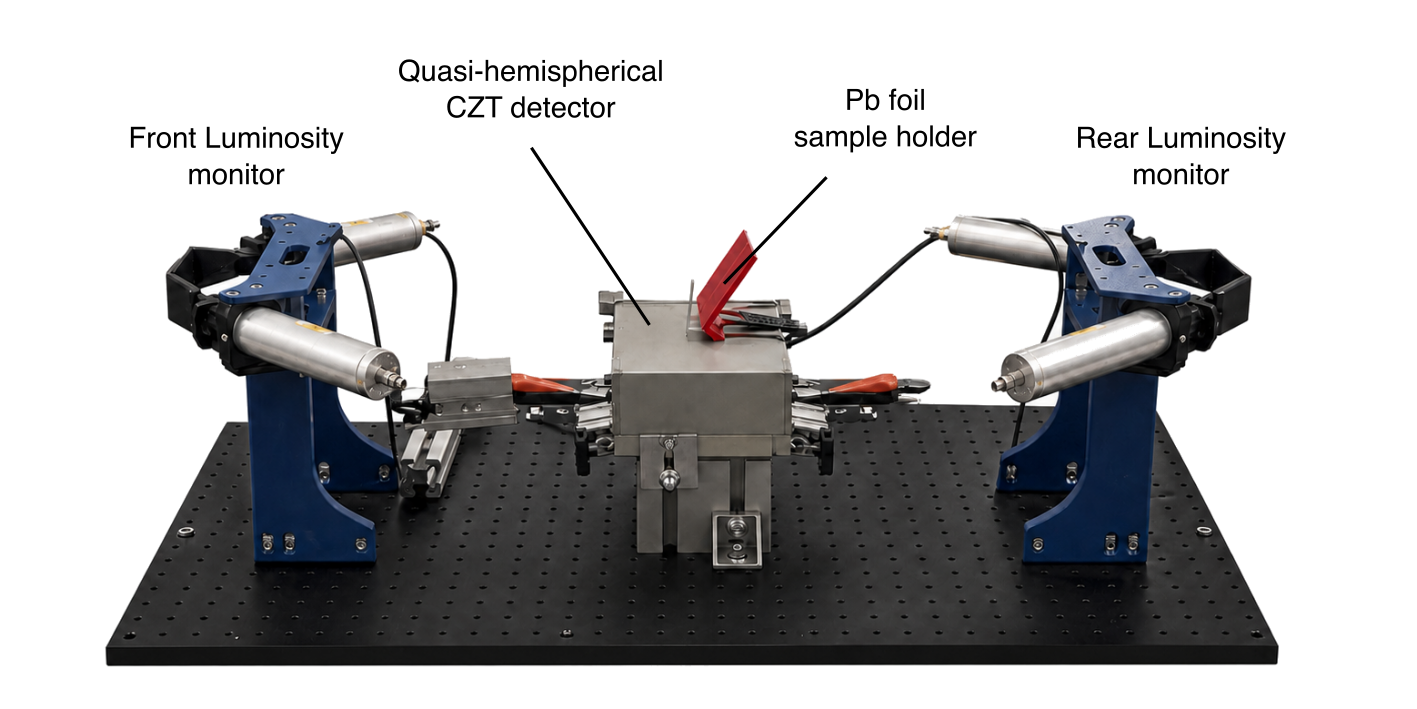}
    \caption{Experimental arrangement used for the measurements at the
    DA$\Phi$NE BTF of INFN-LNF. A \(1~\mathrm{mm}\)-thick Pb target was     placed close to the CZT detector and between the front and rear     luminosity monitors, where each luminosity monitor consisted of a plastic scintillator read out at both ends by photomultiplier tubes.}
    \label{fig:setup}
\end{figure}
%
%
%----------CZT detector and readout----------
The detection system consisted of a quasi-hemispherical CZT crystal with dimensions of \(10\times10\times5~\mathrm{mm^3}\), grown by the Traveling Heater Method by REDLEN Technologies. The crystal and its front-end electronics were enclosed in a thin aluminum housing equipped with a \(0.27~\mathrm{mm}\)-thick entrance window. The quasi-hemispherical electrode geometry was employed to reduce the contribution of holes to the induced signal and to improve the charge-collection properties of the detector.\newline\noindent
%
%----------Experimental geometry and data acquisition----------
Above the CZT detector a \(1~\mathrm{mm}\)-thick Pb target was placed, whose surface was oriented at approximately \(45^{\circ}\) with respect to the incident beam (see Figure \ref{fig:setup}). The detector was positioned close to the target to increase the detection efficiency for the characteristic Pb X-rays. Two luminosity monitors were installed along the beam line, one before and one after the target. Each monitor consisted of a plastic scintillator read out at both ends by photomultiplier tubes and was used to monitor the passage of the electron bunches. The detector signals were digitized using a CAEN VX2740 digitizer with 16-bit resolution and a sampling rate of 125~MS/s. For each trigger, the pulse waveform and its corresponding amplitude were stored for offline analysis. The processed waveform contained 656 samples, corresponding to an approximately \(5~\mu\mathrm{s}\) time interval around the detector pulse. The pulse amplitude was subsequently converted into energy using a linear calibration based on the positions of the Pb \(K\alpha\) and \(K\beta\) lines.\newline\noindent
%----------Waveforms----------
The goal of the PSA is to classify the acquired waveforms. Figure~\ref{fig:waveforms} shows some examples of pulses assigned to the normal and anomalous classes. Normal pulses display a single, regular rising edge followed by a stable plateau (see Figure~\ref{fig:waveforms}a). Anomalous waveforms instead include pulses with multiple steps, pile-up, irregular charge collection, electronic disturbances, or no clearly identifiable physical pulse (see Figure~\ref{fig:waveforms}b).\newline\noindent
%
%----------Pulse-shape features----------
A set of 25 features was extracted from each waveform to quantify its amplitude, temporal development, and degree of regularity. These features included the pulse amplitude and integral, the rise time, the number of peaks in the first derivative, the waveform centroid and root mean square (RMS) width, the curvature, the charge-collection time, different rise-time ratios, and proxies for the depth of interaction (DOI), drift velocity, late charge collection, and ballistic deficit. Three representative features were considered in greater detail for the PSA: the rise time, the number of peaks in the waveform derivative, and the normalized RMS-width proxy, denoted as \doirms. The rise time was calculated as \(t_{90}-t_{10}\), where \(t_{10}\) and \(t_{90}\) are the first times at which the waveform reaches 10\% and 90\% of its maximum amplitude, respectively. The derivative was filtered using a median filter before identifying local maxima, reducing the number of peaks produced by high-frequency fluctuations. The temporal centroid \(t_c\), its weighted RMS width \(\sigma_t\), and the dimensionless \doirms \, feature were calculated within a single definition as
\begin{equation}
t_c =
\frac{\sum_i w_i t_i}{\sum_i w_i},
\qquad
\sigma_t =
\sqrt{
\frac{\sum_i w_i\left(t_i-t_c\right)^2}
     {\sum_i w_i}
},
\qquad
\mathrm{DOI}_{\mathrm{RMS}} =
\frac{\sigma_t}{t_{90}-t_{10}},
\label{eq:doi_rms}
\end{equation}
where \(t_i\) is the time coordinate of the \(i\)-th sample and \(w_i\) is the waveform value. The denominator \(t_{90}-t_{10}\) represents the charge-collection-time proxy used in the analysis. The resulting feature characterizes the temporal spread of the collected charge relative to the pulse rise time. It is therefore sensitive to variations in the charge-collection process, although it should not be interpreted as a direct measurement of the physical interaction depth.\newline\noindent
%
%----------Machine-learning classification----------
Binary classification for the PSA was performed using the XGBoost algorithm. The analysis was implemented in Python using the \verb|scikit-learn| and \verb|xgboost| \cite{Chen_2016_XGBoostScalable} libraries. A subset of 10,000 events was used to train and evaluate the classifier. Initial labels were assigned based on the number of peaks in the first derivative and the rise time. Events were labeled as normal if the first derivative contained exactly one peak and the rise time was within 2$\sigma$ from the mean of 0.2 $\mu$s. The labels were then visually inspected to identify and correct possible false positives and false negatives. The labeled dataset was divided into training and test sets using an 80/20 split. The XGBoost classifier consisted of 300 boosted decision trees with a maximum depth of four, a learning rate of 0.05, a row-subsampling fraction of 0.8, and a feature-subsampling fraction of 0.8 per tree. The classification threshold was selected to maximize the F1-score on the test set.
\section{Preliminary results and discussion}
\begin{figure}[H]
    \centering
    \includegraphics[width=1\linewidth]{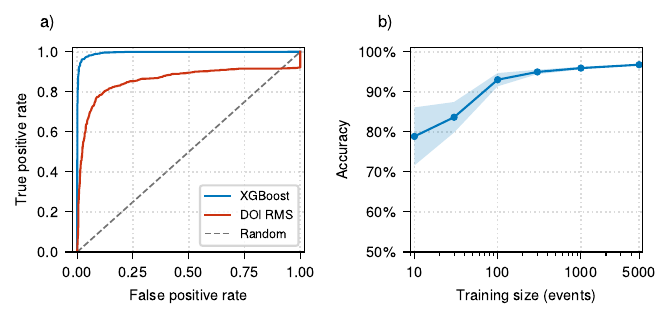}
    \caption{In panel (a) the ROC for XGBoost (blue) and the curve (red) obtained using the \doirms \, feature as the threshold for the classifier. Panel (b) XGBoost classification accuracy as a function of the number of training events. The points and shaded region in panel (b) represent the mean and standard deviation obtained from ten independently drawn stratified training subsets, respectively.}
    \label{fig:ml}
\end{figure}

\begin{figure}[H]
    \centering
    \includegraphics[width=1\linewidth]{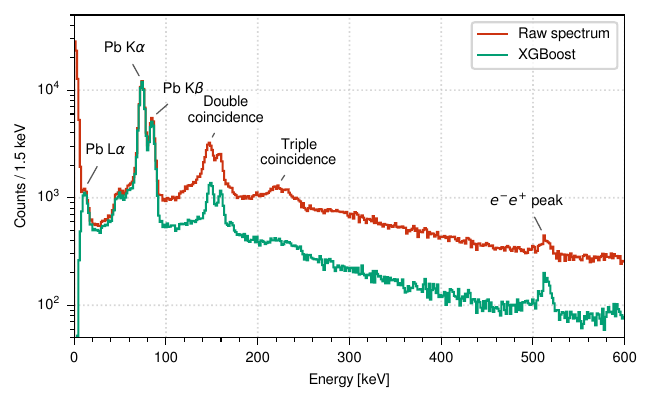}
    \caption{Calibrated CZT energy spectrum before event selection (red) and after retaining events classified as normal by the feature-based XGBoost model (green). The same acquisition exposure and an energy-bin width of 1.5~keV are used for both distributions. The     characteristic Pb lines, the double- and triple-coincidence structures, and the 511-keV electron--positron annihilation peak are indicated.}
    \label{fig:spectrum}
\end{figure}

%
%----------Overview of the results----------
The analysis was organized in three steps. First, the extraction of features from the acquired waveforms. Then, the creation of the training set for training the XGBoost and finally the classification of the entire data with the final selection in the total spectrum.
\begin{figure}[h!]
    \centering
    \includegraphics[width=1\linewidth]{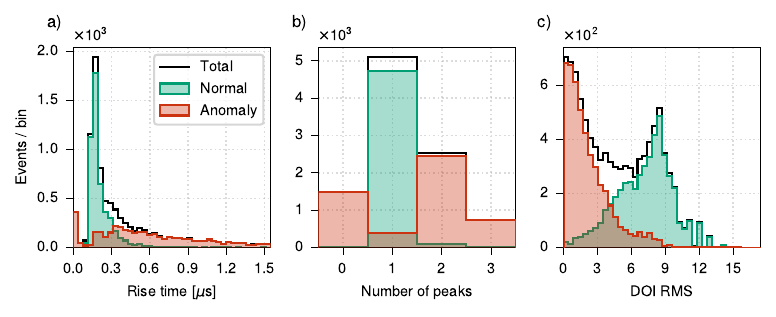}
    \caption{Distributions of three representative pulse-shape features for the manually validated normal (green) and anomalous (red) waveforms; the black histogram represents the complete labeled sample: (a) rise time, (b) number of peaks in the first derivative, and (c) normalized temporal-width proxy \doirms.}
    \label{fig:features}
\end{figure}
%
%
%----------Feature distributions----------
Figure~\ref{fig:features} shows the distributions of the three features: the rise time, the number of peaks in the first derivative, and \doirms \, for the labeled dataset. Each feature provides some discrimination between the two classes, although their distributions partially overlap. As shown in Figure~\ref{fig:features}(a), normal events are concentrated in a relatively narrow rise-time interval around $0.2~\mu\mathrm{s}$. Anomalous events exhibit a broader distribution, with a component close to zero and a long tail extending beyond $1~\mu\mathrm{s}$. Very short rise times are consistent with noise-like pulses, whereas longer rise times may reflect irregular charge collection, pile-up, or temporally separated energy depositions. The number of peaks in the first derivative, shown in Figure~\ref{fig:features}(b), characterizes the structure of the rising edge. Most normal events exhibit a single peak, consistent with a smooth, regular rise. Anomalous events more frequently exhibit either no identifiable peak or multiple peaks, the latter being consistent with multi-step waveforms, pile-up, or noise. Some overlap remains, particularly for low-energy events, in which noise-induced fluctuations along the rising edge can produce additional derivative peaks. The combined feature \doirms \, defined in (\ref{eq:doi_rms}), shown in Figure~\ref{fig:features}(c), exhibits the clearest separation between the two classes. Normal events predominantly populate intermediate and high values, with a maximum around 8--9, whereas anomalous events are concentrated mainly below \doirms $\simeq 4$. Nevertheless, residual overlap limits the discrimination achievable with a single threshold. These distributions therefore motivate a multivariate approach that exploits the complementary information and correlations among the three waveform features using ML methods.
%
%----------Machine-Learning Results----------
%
ML was used to classify waveforms based on pulse-shape features. XGBoost was trained on a subset of the acquired events, with initial labels assigned using feature-based criteria (see Section 2) and subsequently refined through visual inspection. The classification performance obtained using the complete feature set is reported in Figure~\ref{fig:ml}. The receiver operating characteristic (ROC) curves, evaluated on the fixed test sample, are shown in Figure~\ref{fig:ml}(a). XGBoost achieved an area under the curve (AUC) of 0.996 and an accuracy of 96.8\%. To compare ML with respect to using feature selection, we also used the \doirms \, as a threshold, but achieving only an AUC of 0.85, exhibiting the advantage of exploiting the entire feature correlation for classification. We further explore the dependence of the XGBoost accuracy on the number of training events as shown in Figure~\ref{fig:ml}(b). With only a few training events, the mean accuracy is approximately 79\% and exhibits a relatively large variation among the different stratified samples. The accuracy increases to approximately 93\% with a few hundred of training events and approaches 97\% when several thousand labeled waveforms are used. The observed saturation indicates that most of the classification performance can already be obtained with a few hundred labeled events, although a larger training sample improves both accuracy and stability.
%
%
%----------Application to the full dataset----------
After the classification performance had been evaluated on the independent test sample, the final XGBoost model was then applied to 790700 events. This selection accepted 381351 events, corresponding to 48.2\% of the complete dataset.
%
%----------Effect on the energy spectrum----------
Figure~\ref{fig:spectrum} compares the calibrated spectrum of the complete dataset with that obtained after the XGBoost selection. The two distributions use the same acquisition exposure and energy binning and are reported as absolute counts. The unselected spectrum contains the characteristic Pb \(L\alpha\), \(K\alpha\), and \(K\beta\) lines, together with double- and triple-coincidence structures at approximately 150 and 220 keV, respectively and the electron--positron annihilation peak around 511~keV. The XGBoost selection produces a broad reduction of the continuum over the entire investigated energy range while preserving the main characteristic lines. The double- and triple-coincidence regions are suppressed more strongly than the isolated Pb \(K\) lines, consistently with the presence of multi-step or pile-up waveforms among the rejected events. A reduction of the high-energy continuum is also observed around the 511-keV region, where the annihilation peak remains visible after the selection, with improved signal-to-noise ratio. These results demonstrate that the information encoded in the waveform features can be used to improve the relative visibility of spectral structures in the presence of an intense accelerator-induced background.
%
%----------Limitations----------
This preliminary study demonstrates that a few hundred manually labeled waveforms are sufficient to achieve classification accuracy above 90\% relative to the assigned labels and suppress the spectral continuum while preserving the main spectral structures. Further validation across independent acquisitions, detector setups, and beam conditions is needed to establish generalizability. Future developments could employ one-dimensional convolutional neural networks (CNNs) to exploit waveform morphology directly. Integration into the DAQ could enable online event rejection without routinely storing full waveforms, potentially reducing processing time by bypassing explicit feature extraction; this advantage should be verified through latency benchmarks. CNN implementation in FPGA-based acquisition systems provides a precedent for this approach \cite{Astrain_2021_RealTimeImplementation}. ML interpretability methods, like feature importance and SHAP analysis \cite{Lundberg_2020_LocalExplanations} could clarify the features driving the current classifier, while Grad-CAM adapted to one-dimensional signals could identify waveform regions relevant to CNN predictions \cite{Selvaraju_2017_GradCAMVisual}. Finally, Monte Carlo simulations coupled with charge-transport, signal-formation, and electronics-response models could provide physically motivated training labels, reducing reliance on manual annotation, provided that simulated waveforms are validated against measurements \cite{Bettelli_2020_FirstPrinciple,Vicini_2023_OptimizationQuasihemispherical,Yu_2025_PulseShape}.
\section{Conclusions}
%
%----------Main findings----------
This work demonstrates the feasibility of feature-based pulse-shape analysis for background suppression in a quasi-hemispherical CZT detector operated at the DA$\Phi$NE BTF of INFN-LNF. By combining temporal and morphological waveform features, the XGBoost classifier achieved an AUC of 0.996 and an accuracy of 96.8\% on the fixed test sample, with accuracies above 90\% already obtained using a few hundred labeled waveforms. The final model, trained on the complete labeled sample and applied to 790700 events, retained 48.2\% of the data and substantially reduced the spectral continuum while maintaining the visibility of the characteristic Pb lines and the 511-keV annihilation peak. The preferential suppression of the double- and triple-coincidence structures is consistent with the rejection of multi-step and pile-up waveforms. These results establish a proof of concept for machine-learning-assisted PSA and support its further development for CZT spectroscopy in accelerator environments.

%----------Limitations and outlook----------
The classification metrics quantify agreement with morphology-based labels rather than independently verified interaction topologies. Further validation will focus on the energy dependence of signal efficiency and background rejection, including the role of amplitude-related features, and on stability across independent acquisitions and detector conditions. Complementary studies using simulated waveforms will support the physical interpretation of the selected events and the development of the method toward quantitative spectroscopy and online event selection.
\vspace{6pt}

\authorcontributions{Conceptualization, S.M., F.A., C.C. and A.Sc.; methodology, S.M., F.A., L.A. and A.B.; software, S.M., F.A., G.B., M.C. and C.F.; validation, S.M., F.A., L.A., A.B. and A.Sc.; formal analysis, S.M., F.A., F.C. and F.Sg.; investigation, S.M., F.A., M.Ba., D.B., M.Br., A.C., L.D.P., R.D.G., K.D., I.F., G.G., C.G., M.Il., M.Iw., A.K., J.M., P.M., H.O., K.P., F.P., F.Sg., M.Si., D.S., F.Si., M.Sk., A.Sp., K.T., O.V.D. and J.Z.; resources, M.Ba., M.Be., D.B., M.Br., R.D.G., K.D., M.Iw., J.M., P.M., H.O., D.S., F.Si., M.Sk., A.Sp., K.T., A.Z., J.Z. and C.C.; data curation, S.M., F.A., A.C., C.G., M.Il. and O.V.D.; writing-original draft preparation, S.M. and F.A.; writing-review and editing, L.A., M.Ba., M.Be., G.B., D.B., M.Br., A.B., M.C., A.C., F.C., L.D.P., R.D.G., K.D., C.F., I.F., G.G., C.G., M.Il., M.Iw., A.K., J.M., P.M., H.O., K.P., F.P., A.Sc., F.Sg., M.Si., D.S., F.Si., M.Sk., A.Sp., K.T., O.V.D., A.Z., J.Z. and C.C.; visualization, S.M. and F.A.; supervision, C.C., A.Sc., L.A. and A.B.; project administration, S.M., F.A., A.Sc. and C.C.; funding acquisition, C.C., A.Sc., M.Iw. and P.M. All authors have read and agreed to the published version of the manuscript.}

\funding{This work was supported by the INFN KAONNIS project; the Austrian Science Fund (FWF), grants P24756-N20 and P33037-N; the Croatian Science Foundation, project IP-2022-10-3878; the EU STRONG-2020 project, Grant Agreement No. 824093; the EU Horizon 2020 programme under the Marie Skłodowska-Curie Actions, Grant Agreement No. 754496; the Japan Society for the Promotion of Science through JSPS KAKENHI grants JP18H05402 and JP22H04917; the Polish Ministry of Science and Higher Education, grant 7150/E-338/M/2018; the Polish National Agency for Academic Exchange, grant PPN/BIT/2021/1/00037; and COST Action CA24131--ENRICH, supported by COST (European Cooperation in Science and Technology, \url{https://www.cost.eu/}). We gratefully acknowledge the Polish high-performance computing infrastructure PLGrid (HPC Center: ACK Cyfronet AGH) for providing computing facilities and support within computational grant PLG/2025/018524}

\dataavailability{The data supporting the findings of this study are available from the corresponding author upon reasonable request.}

\acknowledgments{We warmly thank Luca Gennaro Foggetta and Eleonora Diociaiuti for their outstanding tailored beam setups and excellent scientific support. We also extend our gratitude to the DA$\Phi$NE operators and LINAC service for maintaining great uptime in BTF injections. We thank C. Capoccia from INFN-LNF and H. Schneider, L. Stohwasser and D. Pristauz-Telsnigg from the Stefan Meyer Institute for their fundamental contributions to the design and construction of the SIDDHARTA-2 setup. Special thanks are extended to Catia Milardi for her continued support and contribution during data taking.}

\conflictsofinterest{The authors declare no conflicts of interest.}

\reftitle{References}

\bibliography{XGRADE}

@article{Abbene_2020_RoomtemperatureXray,
  title = {Room-Temperature {{X-ray}} Response of Cadmium--Zinc--Telluride Pixel Detectors Grown by the Vertical {{Bridgman}} Technique},
  author = {Abbene, L. and Principato, F. and Gerardi, G. and Buttacavoli, A. and Cascio, D. and Bettelli, M. and Amad{\`e}, N. S. and Seller, P. and Veale, M. C. and Fox, O. and Sawhney, K. and Zanettini, S. and Tomarchio, E. and Zappettini, A.},
  year = 2020,
  journal = {Journal of Synchrotron Radiation},
  volume = {27},
  number = {2},
  pages = {319--328},
  publisher = {International Union of Crystallography},
  issn = {1600-5775},
  doi = {10.1107/S1600577519015996},
  langid = {english}
}

@article{Artibani_2026_KaonicCopper,
  title = {Kaonic Copper and Fluorine Absolute Yields Measurement with a {{CZT-based}} Detection System at {{DA$\Phi$NE}}},
  author = {Artibani, Francesco and Manti, Simone and Abbene, Leonardo and Buttacavoli, Antonino and Bettelli, Manuele and Gerardi, Gaetano and Principato, Fabio and Zappettini, Andrea and Bazzi, Massimiliano and Borghi, Giacomo and Bosnar, Damir and Bragadireanu, Mario and Carminati, Marco and Clozza, A and Clozza, Francesco and Del Grande, Raffaele and De Paolis, Luca and Fiorini, Carlo and Fri{\v s}{\v c}i{\'c}, Ivica and Guaraldo, C and Iliescu, M and Iwasaki, Masahiko and Khreptak, Aleksander and Marton, Johann and Moskal, Pawel and Napolitano, Fabrizio and Ohnishi, Hiroaki and Piscicchia, Kristian and Sgaramella, Francesco and Silarski, Micha{\l} and Sirghi, Diana Laura and Sirghi, Florin and Skurzok, Magdalena and Spallone, Antonio and Toho, Kairo and Doce, Oton Vazquez and Zmeskal, Johann and Curceanu, Catalina and Scordo, Alessandro},
  year = 2026,
  journal = {Journal of Physics G: Nuclear and Particle Physics},
  issn = {0954-3899},
  doi = {10.1088/1361-6471/ae9f75},
  langid = {english}
}

@article{Astrain_2021_RealTimeImplementation,
  title = {Real-{{Time Implementation}} of the {{Neutron}}/{{Gamma Discrimination}} in an {{FPGA-Based DAQ MTCA Platform Using}} a {{Convolutional Neural Network}}},
  author = {Astrain, Miguel and Ruiz, Mariano and Stephen, {\relax Adam}. V. and Sarwar, Rashed and Carpe{\~n}o, Antonio and Esquembri, Sergio and Murari, Andrea and Belli, Francesco and Riva, Marco},
  year = 2021,
  month = aug,
  journal = {IEEE Transactions on Nuclear Science},
  volume = {68},
  number = {8},
  pages = {2173--2178},
  issn = {1558-1578},
  doi = {10.1109/TNS.2021.3090670},
  urldate = {2026-09-08}
}

@article{Bettelli_2020_FirstPrinciple,
  title = {A First Principle Method to Simulate the Spectral Response of {{CdZnTe-based X-}} and Gamma-Ray Detectors},
  author = {Bettelli, Manuele and Amad{\`e}, Nicola Sarzi and Calestani, Davide and Garavelli, Bruno and Pozzi, Pietro and Macera, Daniele and Zanotti, Luca and Gonano, Carlo Andrea and Veale, Matthew C. and Zappettini, Andrea},
  year = 2020,
  month = apr,
  journal = {Nuclear Instruments and Methods in Physics Research Section A: Accelerators, Spectrometers, Detectors and Associated Equipment},
  volume = {960},
  pages = {163663},
  issn = {0168-9002},
  doi = {10.1016/j.nima.2020.163663},
  urldate = {2026-09-08}
}

@inproceedings{Chen_2016_XGBoostScalable,
  title = {{{XGBoost}}: A Scalable Tree Boosting System},
  booktitle = {Proceedings of the 22nd {{ACM SIGKDD}} International Conference on Knowledge Discovery and Data Mining},
  author = {Chen, Tianqi and Guestrin, Carlos},
  year = 2016,
  pages = {785--794},
  publisher = {ACM},
  doi = {10.1145/2939672.2939785}
}

@article{Eberth_2008_GeLiDetectors,
  title = {From {{Ge}}({{Li}}) Detectors to Gamma-Ray Tracking Arrays--50 Years of Gamma Spectroscopy with Germanium Detectors},
  author = {Eberth, J. and Simpson, J.},
  year = 2008,
  month = apr,
  journal = {Progress in Particle and Nuclear Physics},
  volume = {60},
  number = {2},
  pages = {283--337},
  issn = {0146-6410},
  doi = {10.1016/j.ppnp.2007.09.001},
  urldate = {2026-09-08}
}

@article{Eisen_1998_CdTeCdZnTe,
  title = {{{CdTe}} and {{CdZnTe}} Materials for Room-Temperature {{X-ray}} and Gamma Ray Detectors},
  author = {Eisen, Y. and Shor, A.},
  year = 1998,
  journal = {Journal of Crystal Growth},
  volume = {184--185},
  pages = {1302--1312},
  issn = {0022-0248},
  doi = {10.1016/S0022-0248(98)80270-4}
}

@inproceedings{Foggetta_2021_ExtendedOperative,
  title = {The Extended Operative Range of the {{LNF LINAC}} and {{BTF}} Facilities},
  booktitle = {12th International Particle Accelerator Conference},
  author = {Foggetta, Luca and others},
  year = 2021,
  doi = {10.18429/JACoW-IPAC2021-THPAB113}
}

@article{Fritts_2014_PulseshapeDiscrimination,
  title = {Pulse-Shape Discrimination of Surface Events in {{CdZnTe}} Detectors for the {{COBRA}} Experiment},
  author = {Fritts, M. and Tebr{\"u}gge, J. and Durst, J. and Ebert, J. and G{\"o}{\ss}ling, C. and G{\"o}pfert, T. and Gehre, D. and Hagner, C. and Heidrich, N. and Homann, M. and K{\"o}ttig, T. and Neddermann, T. and Oldorf, C. and Quante, T. and Rajek, S. and Reinecke, O. and Schulz, O. and Timm, J. and Wonsak, B. and Zuber, K.},
  year = 2014,
  month = jun,
  journal = {Nuclear Instruments and Methods in Physics Research Section A: Accelerators, Spectrometers, Detectors and Associated Equipment},
  volume = {749},
  pages = {27--34},
  issn = {0168-9002},
  doi = {10.1016/j.nima.2014.02.038},
  urldate = {2026-09-08}
}

@article{Gotta_2004_PrecisionSpectroscopy,
  title = {Precision Spectroscopy of Light Exotic Atoms},
  author = {Gotta, D.},
  year = 2004,
  month = mar,
  journal = {Progress in Particle and Nuclear Physics},
  volume = {52},
  number = {1},
  pages = {133--195},
  issn = {0146-6410},
  doi = {10.1016/j.ppnp.2003.09.003},
  urldate = {2026-09-08}
}

@article{Harrison_2013_NUCLEARSPECTROSCOPIC,
  title = {The {{Nuclear Spectroscopic Telescope Array}} ({{NuSTAR}}) High-Energy {{X-ray Mission}}},
  author = {Harrison, Fiona A. and Craig, William W. and Christensen, Finn E. and Hailey, Charles J. and Zhang, William W. and Boggs, Steven E. and Stern, Daniel and Cook, W. Rick and Forster, Karl and Giommi, Paolo and Grefenstette, Brian W. and Kim, Yunjin and Kitaguchi, Takao and Koglin, Jason E. and Madsen, Kristin K. and Mao, Peter H. and Miyasaka, Hiromasa and Mori, Kaya and Perri, Matteo and Pivovaroff, Michael J. and Puccetti, Simonetta and Rana, Vikram R. and Westergaard, Niels J. and Willis, Jason and Zoglauer, Andreas and An, Hongjun and Bachetti, Matteo and Barri{\`e}re, Nicolas M. and Bellm, Eric C. and Bhalerao, Varun and Brejnholt, Nicolai F. and Fuerst, Felix and Liebe, Carl C. and Markwardt, Craig B. and Nynka, Melania and Vogel, Julia K. and Walton, Dominic J. and Wik, Daniel R. and Alexander, David M. and Cominsky, Lynn R. and Hornschemeier, Ann E. and Hornstrup, Allan and Kaspi, Victoria M. and Madejski, Greg M. and Matt, Giorgio and Molendi, Silvano and Smith, David M. and Tomsick, John A. and Ajello, Marco and Ballantyne, David R. and Balokovi{\'c}, Mislav and Barret, Didier and Bauer, Franz E. and Blandford, Roger D. and Brandt, W. Niel and Brenneman, Laura W. and Chiang, James and Chakrabarty, Deepto and Chenevez, Jerome and Comastri, Andrea and Dufour, Francois and Elvis, Martin and Fabian, Andrew C. and Farrah, Duncan and Fryer, Chris L. and Gotthelf, Eric V. and Grindlay, Jonathan E. and Helfand, David J. and Krivonos, Roman and Meier, David L. and Miller, Jon M. and Natalucci, Lorenzo and Ogle, Patrick and Ofek, Eran O. and Ptak, Andrew and Reynolds, Stephen P. and Rigby, Jane R. and Tagliaferri, Gianpiero and Thorsett, Stephen E. and Treister, Ezequiel and Urry, C. Megan},
  year = 2013,
  journal = {The Astrophysical Journal},
  volume = {770},
  number = {2},
  pages = {103},
  publisher = {The American Astronomical Society},
  issn = {0004-637X},
  doi = {10.1088/0004-637X/770/2/103},
  langid = {english}
}

@article{Holl_2019_DeepLearninga,
  title = {Deep Learning Based Pulse Shape Discrimination for Germanium Detectors},
  author = {Holl, P. and Hauertmann, L. and Majorovits, B. and Schulz, O. and Schuster, M. and Zsigmond, A. J.},
  year = 2019,
  month = may,
  journal = {The European Physical Journal C},
  volume = {79},
  number = {6},
  pages = {450},
  issn = {1434-6052},
  doi = {10.1140/epjc/s10052-019-6869-2},
  urldate = {2026-09-08},
  langid = {english}
}

@book{Iniewski_2024_CdTeCdZnTe,
  title = {{{CdTe}} and {{CdZnTe Materials}}: {{Material Properties}} and {{Applications}}},
  shorttitle = {{{CdTe}} and {{CdZnTe Materials}}},
  author = {Iniewski, Kris},
  year = 2024,
  publisher = {Springer Nature},
  googlebooks = {JpkbEQAAQBAJ},
  isbn = {978-3-031-64521-1},
  langid = {english}
}

@article{Lundberg_2020_LocalExplanations,
  title = {From Local Explanations to Global Understanding with Explainable {{AI}} for Trees},
  author = {Lundberg, Scott M. and Erion, Gabriel and Chen, Hugh and DeGrave, Alex and Prutkin, Jordan M. and Nair, Bala and Katz, Ronit and Himmelfarb, Jonathan and Bansal, Nisha and Lee, Su-In},
  year = 2020,
  month = jan,
  journal = {Nature Machine Intelligence},
  volume = {2},
  number = {1},
  pages = {56--67},
  publisher = {Nature Publishing Group},
  issn = {2522-5839},
  doi = {10.1038/s42256-019-0138-9},
  urldate = {2026-09-09},
  copyright = {2020 The Author(s), under exclusive licence to Springer Nature Limited},
  langid = {english}
}

@article{Manti_2025_EXKALIBURKaonic,
  title = {{{EXKALIBUR}}: {{Towards}} a {{Kaonic Atoms Periodic Table}} to {{Test Fundamental Interactions}}},
  shorttitle = {{{EXKALIBUR}}},
  author = {Manti, S. and Khreptak, A. and Marton, J. and Moskal, P. and Ohnishi, H. and Pischicchia, K. and Principato, F. and Scordo, A. and Sgaramella, F. and Silarski, M. and Sirgh, D. and Sirghi, F. and Skurzok, M. and Spallone, A. and Toho, K. and Toscano, L. and Doce, O. Vazquez and Iwasaki, M. and Indelicato, P. and Iliescu, M. and Abbene, L. and Artibani, F. and Bazzi, M. and Borghi, G. and Bosnar, D. and Bragadireanu, M. and Buttacavoli, A. and Carminati, M. and Clozza, F. and Clozza, A. and Paolis, L. De and Grande, R. Del and Dulski, K. and Fabbietti, L. and Fiorini, C. and Fri{\v s}{\v c}i{\'c}, I. and Curceanu, C.},
  year = 2025,
  month = dec,
  journal = {Acta Physica Polonica A},
  volume = {148},
  number = {6},
  pages = {S89-S89},
  issn = {1898-794X},
  doi = {10.12693/APhysPolA.148.S89},
  urldate = {2026-09-07},
  copyright = {Copyright (c) 2025 Acta Physica Polonica A},
  langid = {english}
}

@article{Manti_2026_MachinelearningenhancedEvent,
  title = {Machine-Learning-Enhanced Event Selection for {{BEGe}} Detectors in the {{VIP}} Experiment},
  author = {Manti, S. and Yip, S.H. and Bazzi, M. and Bortolotti, N. and Bragadireanu, M. and Carnevali, I. and Clozza, A. and De Paolis, L. and Del Grande, R. and Guaraldo, C. and Iliescu, M. and Laubenstein, M. and Marton, J. and Nola, F. and Piscicchia, K. and Porcelli, A. and Scordo, A. and Sgaramella, F. and Sirghi, D. and Sirghi, F. and Zmeskal, J. and Curceanu, C.},
  year = 2026,
  month = jan,
  journal = {Journal of Instrumentation},
  volume = {21},
  number = {01},
  pages = {P01036},
  publisher = {IOP Publishing},
  issn = {1748-0221},
  doi = {10.1088/1748-0221/21/01/P01036},
  urldate = {2026-09-07},
  langid = {english}
}

@article{McGrath_2010_DetectingMultihit,
  title = {Detecting Multi-Hit Events in a {{CdZnTe}} Coplanar Grid Detector Using Pulse Shape Analysis: {{A}} Method for Improving Background Rejection in the {{COBRA}} 0{$\nu\beta\beta$} Experiment},
  shorttitle = {Detecting Multi-Hit Events in a {{CdZnTe}} Coplanar Grid Detector Using Pulse Shape Analysis},
  author = {McGrath, J. and Fulton, B. R. and Joshi, P. and Davies, P. and Muenstermann, D. and Schulz, O. and Zuber, K. and Freer, M.},
  year = 2010,
  month = mar,
  journal = {Nuclear Instruments and Methods in Physics Research Section A: Accelerators, Spectrometers, Detectors and Associated Equipment},
  volume = {615},
  number = {1},
  pages = {57--61},
  issn = {0168-9002},
  doi = {10.1016/j.nima.2010.01.025},
  urldate = {2026-09-08}
}

@article{Mele_2023_AdvancesHighEnergyResolution,
  title = {Advances in {{High-Energy-Resolution CdZnTe Linear Array Pixel Detectors}} with {{Fast}} and {{Low Noise Readout Electronics}}},
  author = {Mele, Filippo and Quercia, Jacopo and Abbene, Leonardo and Benassi, Giacomo and Bettelli, Manuele and Buttacavoli, Antonino and Principato, Fabio and Zappettini, Andrea and Bertuccio, Giuseppe},
  year = 2023,
  month = feb,
  journal = {Sensors},
  volume = {23},
  number = {4},
  publisher = {Multidisciplinary Digital Publishing Institute},
  issn = {1424-8220},
  doi = {10.3390/s23042167},
  urldate = {2026-07-08},
  copyright = {http://creativecommons.org/licenses/by/3.0/},
  langid = {english}
}

@inproceedings{Milardi_2018_PreparationActivitya,
  title = {Preparation {{Activity}} for the {{SIDDHARTA-2 Run}} at {{DA$\Phi$NE}}},
  booktitle = {9th International Particle Accelerator Conference ({{IPAC2018}})},
  author = {Milardi, Catia and Alesini, David and Bini, Simone and {Blanco-Garc{\'i}a}, Oscar and Boscolo, Manuela and Buonomo, Bruno and Cantarella, Sergio and Caschera, Salvatore and Castorina, Giovanni and Chavanne, Joel and others},
  year = 2018,
  pages = {334--337},
  publisher = {JACOW Publishing, Geneva, Switzerland},
  doi = {10.18429/JACoW-IPAC2018-MOPMF088},
  isbn = {978-3-95450-184-7}
}

@inproceedings{Milardi_2021_DAFNECommissioning,
  title = {{{DA$\Phi$NE Commissioning}} for the {{SIDDHARTA-2 Experiment}}},
  booktitle = {12th International Particle Accelerator Conference ({{IPAC2021}})},
  author = {Milardi, Catia and Alesini, David and {Blanco-Garc{\'i}a}, Oscar and Boscolo, Manuela and Buonomo, Bruno and Cantarella, Sergio and Chavanne, Joel and D'Uffizi, Alessandro and De Santis, Antonio and Di Giulio, Claudio and others},
  year = 2021,
  pages = {1322--1325},
  publisher = {JACOW Publishing, Geneva, Switzerland},
  doi = {10.18429/JACoW-IPAC2021-TUPAB001},
  isbn = {978-3-95450-214-1}
}

@inproceedings{Milardi_2024_DAFNEOperation,
  title = {{{DA$\Phi$NE}} Operation Strategy for the Observation of Kaonic Deuterium},
  booktitle = {15th International Particle Accelerator Conference ({{IPAC2024}})},
  author = {Milardi, Catia and Alesini, David and {Blanco-Garc{\'i}a}, Oscar and Boscolo, Manuela and Buonomo, Bruno and Cantarella, Sergio and Chavanne, Joel and D'Uffizi, Alessandro and De Santis, Antonio and Di Giulio, Claudio and others},
  year = 2024,
  pages = {2504--2507},
  publisher = {JACoW Publishing, Geneva, Switzerland},
  doi = {10.18429/JACoW-IPAC2024-WEPR17},
  isbn = {978-3-95450-247-9}
}

@article{Nakhostin_2013_ApplicationPulseshape,
  title = {Application of Pulse-Shape Discrimination to Coplanar {{CdZnTe}} Detectors},
  author = {Nakhostin, M. and Podolyak, {\relax Zs}. and Sellin, P. J.},
  year = 2013,
  month = nov,
  journal = {Nuclear Instruments and Methods in Physics Research Section A: Accelerators, Spectrometers, Detectors and Associated Equipment},
  volume = {729},
  pages = {541--545},
  issn = {0168-9002},
  doi = {10.1016/j.nima.2013.07.073},
  urldate = {2026-09-08}
}

@article{Renner_2019_ChallengesXray,
  title = {Challenges of X-Ray Spectroscopy in Investigations of Matter under Extreme Conditions},
  author = {Renner, O. and Rosmej, F. B.},
  year = 2019,
  journal = {Matter and Radiation at Extremes},
  volume = {4},
  number = {2},
  pages = {024201},
  issn = {2468-2047},
  doi = {10.1063/1.5086344}
}

@article{Scordo_2024_CdZnTeDetectors,
  title = {{{CdZnTe}} Detectors Tested at the {{DA$\Phi$NE}} Collider for Future Kaonic Atoms Measurements},
  author = {Scordo, A. and Abbene, L. and Artibani, F. and Bazzi, M. and Bettelli, M. and Bosnar, D. and Borghi, G. and Bragadireanu, M. and Buttacavoli, A. and Cargnelli, M. and Carminati, M. and Clozza, A. and Clozza, F. and De Paolis, L. and Deda, G. and Del Grande, R. and Fabbietti, L. and Fiorini, C. and Fri{\v s}{\v c}i{\'c}, I. and Guaraldo, C. and Iliescu, M. and Iwasaki, M. and Khreptak, A. and Manti, S. and Marton, J. and Moskal, P. and Napolitano, F. and Nied{\'z}wiecki, S. and Ohnishi, H. and Piscicchia, K. and Principato, F. and Sada, Y. and Sgaramella, F. and Silarski, M. and Sirghi, D. L. and Sirghi, F. and Skurzok, M. and Spallone, A. and Toho, K. and T{\"u}chler, M. and Yoshida, C. and Zappettini, A. and Zmeskal, J. and Curceanu, C.},
  year = 2024,
  journal = {Nuclear Instruments and Methods in Physics Research Section A: Accelerators, Spectrometers, Detectors and Associated Equipment},
  volume = {1060},
  pages = {169060},
  issn = {0168-9002},
  doi = {10.1016/j.nima.2023.169060}
}

@inproceedings{Selvaraju_2017_GradCAMVisual,
  title = {Grad-{{CAM}}: {{Visual Explanations From Deep Networks}} via {{Gradient-Based Localization}}},
  shorttitle = {Grad-{{CAM}}},
  booktitle = {Proceedings of the {{IEEE International Conference}} on {{Computer Vision}}},
  author = {Selvaraju, Ramprasaath R. and Cogswell, Michael and Das, Abhishek and Vedantam, Ramakrishna and Parikh, Devi and Batra, Dhruv},
  year = 2017,
  pages = {618--626},
  urldate = {2026-09-08}
}

@article{Sirghi_2024_SIDDHARTA2Apparatusb,
  title = {{{SIDDHARTA-2}} Apparatus for Kaonic Atoms Research},
  author = {Sirghi, F. and others},
  year = 2024,
  journal = {Journal of Instrumentation},
  volume = {19},
  number = {11},
  pages = {P11006},
  doi = {10.1088/1748-0221/19/11/P11006}
}

@article{Taguchi_2013_Vision20,
  title = {Vision 20/20: {{Single-Photon-Counting X-ray Detectors}} in Medical Imaging},
  shorttitle = {Vision 20/20},
  author = {Taguchi, Katsuyuki and Iwanczyk, Jan S.},
  year = 2013,
  journal = {Medical Physics},
  volume = {40},
  number = {10},
  pages = {100901},
  issn = {2473-4209},
  doi = {10.1118/1.4820371},
  langid = {english}
}

@article{Vicini_2023_OptimizationQuasihemispherical,
  title = {Optimization of Quasi-Hemispherical {{CdZnTe}} Detectors by Means of First Principles Simulation},
  author = {Vicini, Valentina and Zanettini, Silvia and Sarzi Amad{\`e}, Nicola and Grill, Roman and Zambelli, Nicola and Calestani, Davide and Zappettini, Andrea and Abbene, Leonardo and Bettelli, Manuele},
  year = 2023,
  month = feb,
  journal = {Scientific Reports},
  volume = {13},
  number = {1},
  pages = {3212},
  publisher = {Nature Publishing Group},
  issn = {2045-2322},
  doi = {10.1038/s41598-023-30181-2},
  urldate = {2026-09-14},
  copyright = {2023 The Author(s)},
  langid = {english}
}

@article{Yu_2025_PulseShape,
  title = {Pulse Shape Discrimination Using a Convolutional Neural Network in an Inverted Coaxial Point Contact {{HPGe}} Detector},
  author = {Yu, Siyuan and He, Li and Li, Yulan and Tian, Yang and Yang, Mingxin},
  year = 2025,
  month = mar,
  journal = {Nuclear Instruments and Methods in Physics Research Section A: Accelerators, Spectrometers, Detectors and Associated Equipment},
  volume = {1072},
  pages = {170219},
  issn = {0168-9002},
  doi = {10.1016/j.nima.2025.170219},
  urldate = {2026-09-08}
}

\end{document}